\documentstyle [12pt] {article}

\newcommand {\be} {\begin{equation}}
\newcommand {\bea} {\begin{eqnarray}}
\newcommand {\ee} {\end{equation}}
\newcommand {\eea} {\end{eqnarray}}
\newcommand {\bi} {\bibitem}
\newcommand {\r} {\vec{r}}
\newcommand {\x} {\vec{x}}
\newcommand {\y} {\vec{y}}
\newcommand {\p} {\psi}
\newcommand {\la} {\lambda}
\newcommand {\gd} {g^{\dagger}}
\newcommand {\ex} { e^{ l \beta h s}}
\newcommand {\exj} { e^{ j \beta h s}}
\newcommand {\exm} { e^{ - j \beta h s}}
\newcommand {\exs} { e^{ - j \beta h}}
\newcommand {\ie} {{\it i.~e.}}
\newcommand {\eg} {{\it e.~g.}}

\begin{document}
\title {Theory of continuum percolation I. General formalism }
\author{ Alon Drory}
\date{ \it Dipartimento di Fisica, Universit\'a {\sl La Sapienza}\\
        Piazzale Aldo Moro 2, Roma 00187, Italy}
\maketitle

\begin{abstract}
The theoretical basis of continuum percolation has changed greatly since its
beginning as little more than an analogy with lattice systems. Nevertheless, 
there is yet no comprehensive theory of this field. A basis for such a theory 
is provided here with the introduction of the Potts fluid, a system of 
interacting $s$-state spins which are free to move in the continuum. In the 
$s \to 1$ limit, the Potts magnetization, susceptibility and correlation 
functions are directly related to the percolation probability, the mean cluster
size and the pair-connectedness, respectively. Through the Hamiltonian
formulation  of the Potts fluid, the standard methods of statistical mechanics
can therefore be used in the continuum percolation problem.
\end{abstract}

\newpage

\section{Introduction}
\setcounter{equation}{0} 

The theoretical basis of continuum percolation \cite{bal} changed drastically 
during the 1970's. In their seminal 1970 paper, Scher and Zallen \cite{scher} 
considered continuum percolation as an extension of lattice models. Noting 
that the critical volume $\phi_c$ is approximately universal, \ie, essentially 
independent of the lattice structure, they suggested it might be carried over 
into continuum systems.

Such a view of the continuum as the representation of what is 
lattice-independent sits well with the ideas of the renormalization group where
the continuum often represents merely a scale at which the underlying lattice 
blurs and vanishes.

Yet continuum percolation turned out to be much richer than suggested by this 
conception. In the first extensive simulations of continuum systems, Pike and 
Seager \cite{pike} found that while the critical volume concept works fairly 
well for spherical objects, it fails completely when the system contains 
elongated bodies. This and similar works \cite{binen} prompted Balberg et al.
\cite{balberg} to suggest that rather than the critical volume, the proper 
universal quantity might be a critical {\it excluded} volume. This concept,
however, cannot be reduced to some underlying lattice structure. Instead, it
requires considering the continuum as fundamental rather than a large scale
approximation. 

Fruitful as the excluded volume concept was, it nevertheless had clear 
limitations, since the critical excluded volume is not in fact a truly 
universal quantity. The shape of the objects in the system  clearly influenced
the critical point -- \ie, the critical density, $\rho_c$, at which an infinite
cluster first appears -- in a complicated way.

In 1977, Haan and Zwanzig \cite{haan} managed to push to its limits the analogy
between continuum and lattice percolation. Relying heavily on graph theory, 
they managed to expand the mean cluster size, $S$, in powers of the density.
From this, they obtained values for the critical density which were in good
quantitative agreement with the results of simulations. However, this line of 
developments was reaching its limits as it became clear that continuum 
percolation could be a complex interplay involving not only the shape of 
objects -- which determined the binding criterion -- but also possible
interactions between these objects \cite{int}. This new element was 
beyond accommodation into the analogy with lattice models.

The breakthrough which gave continuum percolation an independent theoretical 
basis came in two papers by Coniglio et al. \cite{coniglio}. Building on 
previous work by Hill \cite{hill}, these researchers showed that the mean 
cluster size could be related to a pair-connectedness function which had an 
expansion in powers of the density. This expansion included naturally the
interactions. Originally, Coniglio et al. developed this expansion in the 
context of physical clustering in a gas, but it was soon extended by analogy to
other systems \cite{des, bug}. Although the results were mainly qualitative at 
first, some recent calculations have obtained quantitatively good agreement 
with simulations \cite{alon}.

In spite of these successes, the theoretical basis of continuum percolation is
still incomplete. The theory of Coniglio et al. suffers from two essential 
problems. The first is formal. The fundamental quantity in this theory is the 
pair-connectedness function, yet the only mathematical definition of it is in
terms of a power expansion. This is an unsatisfactory situation. Some 
non-perturbative scheme must underly such a result, but it is absent from the
work of Coniglio et al.. Another unsatisfactory aspect is that the extension
of this work to various percolation systems has to be done by analogy and 
reasonable assumption. While this is not a serious problem, one would like to 
see a more formal basis for such uses. The second main problem is that the
theory of Coniglio et al. does not say anything about the order parameter,
the percolation probability. This problem is related to the first because
perturbative expansions are limited to the region of densities below the 
critical point, while the order parameter is non-trivial only at densities 
above this point. Furthermore, the percolation probability is not obtainable 
from the pair-connectedness, and requires therefore a new approach.

Such limitations also prevent one from approaching some fundamental questions.
First among these, probably, is related to the continuum percolation 
universality class. Computer simulations suggest that continuum percolation 
belongs to the universality class of lattice percolation. One would like to
understand this from the theoretical point of view, as well as to assess 
whether this is true for all possible interactions and binding criteria. 
Computer simulations are inadequate to address this problem since they are
greatly complicated by the presence of interactions. Furthermore, obtaining
critical exponents from a simulation is a notoriously delicate procedure, and
one cannot claim yet that the universality class of continuum percolation is
known with certainty.

The present series of papers is meant to be the basis for a more comprehensive
theory of continuum percolation, which is based on an extension of the classic
mapping between the Potts model and lattice percolation which was invented by
Kasteleyn and Fortuin \cite{kast}. The present extension includes an 
off-lattice version of the Potts model which I have called the Potts fluid. The
$s$-state Potts fluid is introduced in section 2. I show then that, as in the
lattice case, all statistical averages in the percolation model may be 
expressed as averages in the Potts fluid. In section 3, this mapping is applied
to the Potts magnetization and susceptibility. The limit $s \to 1$ of these 
quantities yields the percolation probability, $P(\rho)$, and the mean cluster
size, $S$, respectively. Section 4 is concerned with the Potts correlation 
functions. The pair-connectedness is shown to be directly related to the Potts
pair-correlation function. Section 5 sums up the results.

\section {General formalism}
\setcounter{equation}{0} 

By ``continuum percolation'' I always mean a system of particles interacting 
through a pair potential $ v(\r_i, \r_j)$ and obeying classical statistical
mechanics. In addition to their interaction, the particles possess another 
property, ``the connectivity'', determined by a probability
function $ p(\r_i,\r_j)$ which is the probability that two particles
located at $\r_i$ and $\r_j$ are bound, or connected, to each other. A {\it 
cluster} is a group of such connected particles. Let us also define 
$ q(\r_i,\r_j) = 1 - p(\r_i,\r_j)$ as the complementary function. 

Note that we do not assume any ``locking'' mechanism: bound particles do not
remain glued together thereafter. Indeed, the system does not evolve with time 
at all. The point of view adopted here is that of equilibrium statistical
mechanics, where all properties are derived from an ensemble of ``snapshot''
configurations. As a result, the properties of connectivity and interaction are
arranged hierarchically. The interaction alone determines the configuration of
the particles, which in turn helps determine the connectivity, but the 
connectivity does not in its turn influence the configuration.

Often, as in systems of permeable objects \cite{pike}, $ p(\r_i,\r_j)$ takes 
only two values, 0 or 1, but no such restrictions are assumed here. Thus, the 
connectivity state need not be uniquely determined by the geometrical 
configuration of the particles. Only its probability distribution is fully
determined through the function $ p(\r_i,\r_j)$.

The formalism presented here is an extension of the Kasteleyn-Fortuin mapping 
between lattice percolation and the Potts model \cite{kast}. For the continuum 
case I will define an extension of the Potts model hereafter called the
{\it Potts fluid}.

The $s$-state Potts fluid is a system of N ``spins'' $\{\la_i\}_{i=1}^N$ each 
having $s$ possible states, and obeying classical statistical mechanics. Each
spin has a position $\r_i$ in the continuum and the spins interact with each 
other through a spin-dependent pair potential $V(\r_i,\la_i;\r_j,\la_j)$, such
that
\be
V(\r_i,\la_i;\r_j,\la_j) \equiv V(i,j) = \left\{ \begin{array}{r @{\quad 
         \mbox{if} \quad}l}
   U(\r_i,\r_j) & \la_i =\la_j \\ W(\r_i,\r_j) & \la_i \neq\la_j 
   \end{array} \right.    \label {poten}
\ee
where $U$ and $W$ are arbitrary functions.

The spins may also couple to an external field $h(\r)$ which tends to 
align them in some given state. If we arbitrarily denote this state as ``1'' ,
then the interaction Hamiltonian is
\be
H_{int}=  - \sum_{i=1}^N \,\p (\la_i) h(\r_i)
\ee
where
\be
\p (\la) = \left\{ \begin{array}{r @{\quad \mbox{if} \quad} l}
            s - 1 & \la = 1 \\ -1 & \la \ne 1 \end{array} \right. \label{psi}
\ee

For conciseness, two spins in the same state will be said to be {\it parallel}
to each other (though the state need not actually correspond to any spatial 
directions). Two spins in different states will be called non-parallel.

As in all classical systems \cite{hansen}, the dependence on the momenta can be
factored out so that all statistical averages depend on the configuration
integral
\be
Z = \frac{1}{N!} \sum_{\{\la_m\}} \int \, d\r_1 \cdots d\r_N \,
\exp \left[ -\beta \sum_{i>j} V(i,j) + \beta \sum_{i=1}^N h(i) \p (\la_i) 
\right] \label{conf}
\ee
where the sum $\sum_{\{\la_m\}}$ is over all spin configurations, and 
$\beta = 1 / kT$ is the inverse temperature as usual.

Now, any continuum percolation model defined by $v(i,j)$ and $p(i,j)$ can be 
mapped onto an appropriate Potts fluid model with a pair-spin interaction 
defined by
\bea
U(i,j) & = & v(i,j) \nonumber\\
\exp \left[- \beta W(i,j) \right] & = & q(i,j) \exp \left[ - \beta v(i,j) 
\right]
\label{map}
\eea

This mapping relates statistical averages of the Potts fluid to statistical
averages of the percolation model. The most fundamental relation is obtained 
for the Potts fluid configuration integral Eq.~(\ref{conf}). However, because 
the following derivation can be a little confusing, let us anticipate the final
result. The formal mapping Eq.~(\ref{map}) induces a geometrical mapping 
between spin configurations and connectivity states. To every particle in the 
percolation system we assign a spin in such a way that if two objects belong to
the same cluster in the percolation model, they are assigned the same spin 
(i.e, their spins are parallel) in the corresponding Potts fluid spin 
configuration. However, the actual value of this common spin is selected at 
random. Therefore the mapping from the percolation model to the Potts fluid is 
one to many. Any 
connectivity state corresponds to several spin configurations, which differ 
from one another by the actual value of the spin assigned in common to all the 
particles in a given cluster. Because of the one to many character of this 
mapping, the status of two parallel spins differs radically from the status of 
two non-parallel spins. If two spins are non-parallel, they {\it must} belong 
to different clusters in the percolation picture. However, if two spins are 
parallel {\it no conclusion} may be drawn as to their connectivity state. They
might be parallel because they belong to the same cluster. But they might also
belong to different clusters and have been assigned the same spin by chance.
Since the common spin of every particle in a cluster is assigned randomly, it
is quite probable that two unrelated clusters end up with the same spin. This
difference makes ``disconnectedness'' a more basic property than connectedness
in the sense that given a spin configuration, one can never deduce with
certainty that two particles belong to the same cluster in the corresponding
percolation model. However, one  may be able to deduce that two particle 
{\it do not} belong to the same cluster.

Let us now see how this geometrical mapping arises naturally out of the formal
mapping Eq.~(\ref{map}). For simplicity, let us assume temporarily that the 
external field vanishes, \ie, $h(\r) = 0$. Introduce the notation
\bea
Z &=& \frac{1}{N!}  \int d\r_1 \cdots d\r_N \, Q(\r_1, \ldots, \r_N) 
\nonumber \\
Q(\r_1, \ldots,\r_N) &=& \sum_{\{\la_m\}}\exp \left[ -\beta \sum_{i>j} V(\r_i,
\r_j) \right] \label{Q}
\eea
In the expression for $Q(1, \ldots, N)$, let us separate all possible 
configurations ${\la_m}$ into those where $\la_1 = \la_2$ and the rest
\bea
Q &=& e^{-\beta U(1,2)} \sum_{\left\{{\la_m \atop  \la_1 = \la_2}\right\}} 
\exp \left[ -\beta \sum_{(i,j)-(1,2)}V(i,j) \right] \nonumber \\
&+ & e^{-\beta W(1,2)} \sum_{\left\{{\la_m \atop  \la_1 \ne \la_2}\right\}} 
\exp \left[ -\beta \sum_{(i,j)-(1,2)}V(i,j) \right]
\label{st1}
\eea
where $\sum_{(i,j)-(1,2)}$ means a summation over all pairs $(i,j)$ except the
pair $(1,2)$. In the second term on the l.h.s of Eq.~(\ref{st1}) we can now 
rewrite
\bea
\sum_{\left\{{\la_m \atop  \la_1 \ne \la_2}\right\}} \exp \left[ -\beta
\sum_{(i,j)-(1,2)}V(i,j)\right] &=& \sum_{\{\la_m\}} \exp 
\left[-\beta \sum_{(i,j)-(1,2)}V(i,j)\right] \nonumber \\
&- & \! \sum_{\left\{{\la_m \atop \la_1 = \la_2}\right\}} \exp \left[-\beta 
\sum_{(i,j)-(1,2)}V(i,j) \right] \nonumber \\
\label{st2}
\eea
where the sum $\sum_{\{\la_m\}}$ is now performed over all spin configurations
without constraints. Eq.~(\ref{st1}) now becomes
\bea
Q &=& \left[ e^{-\beta U(1,2)}- e^{\beta W(1,2)} \right] \sum_{\left\{{\la_m 
\atop  \la_1 = \la_2}\right\}}\exp \left[ -\beta \sum_{(i,j)-(1,2)}V(i,j) 
\right] \nonumber \\
&+& e^{-\beta W(1,2)} \sum_{\{\la_m\}} \exp \left[ -\beta \sum_{(i,j)
-(1,2)}V(i,j)\right]    
\label{st3}
\eea
The mapping Eq.~(\ref{map}) then implies that
\bea
Q &=& p(1,2)e^{-\beta v(1,2)} \sum_{\left\{{\la_m \atop  \la_1 = \la_2}
\right\}} \exp \left[ -\beta \sum_{(i,j)-(1,2)}V(i,j)\right] \nonumber \\
&+ & q(1,2)e^{-\beta v(1,2)} \sum_{\{\la_m\}} 
\exp \left[ -\beta \sum_{(i,j)-(1,2)}V(i,j) \right]
\label{st4}
\eea
where we used the fact that $p(1,2) = 1 - q(1,2)$. Repeating this procedure for
the pair $(1,3)$, we obtain
\bea
Q &=& p(1,2)\, p(1,3)e^{- \beta[v(1,2) + v(1,3)]} 
\sum_{\left\{{\la_m \atop \la_1 = \la_2 =\la_3}\right\}} \exp \left[ -\beta 
\sum_{(i,j)-(1,2)-(1,3)}V(i,j) \right] \nonumber \\
& +&  p(1,2)\, q(1,3)e^{-\beta[ v(1,2) + v(1,3)]} \sum_{\left\{{\la_m}\atop 
{\la_1 =\la_2}\right\}} \exp \left[ -\beta \sum_{(i,j)-(1,2)-(1,3)}V(i,j) 
\right] \nonumber \\
& +& q(1,2) \, p(1,3)e^{-\beta[v(1,2) + v(1,3)]} \sum_{\left\{{\la_m} \atop
{\la_1 = \la_3}\right\}} \exp \left[ -\beta \sum_{(i,j)-(1,2)-(1,3)}V(i,j)
\right] \nonumber \\
& +& q(1,2)\, q(1,3)e^{-\beta[v(1,2) + v(1,3)]} \sum_{\{\la_m\}} \exp 
\left[-\beta \sum_{(i,j)-(1,2)}V(i,j) \right]
\label{st5}
\eea

One sees therefore how the geometric mapping comes about. Every sum is 
performed over a subset of spin configurations determined by a set of {\it
constraints} of the form $\la_1=\la_2, \ldots$. All constraints are equalities
between spin pairs. There are no constraints forcing two spins to be 
non-parallel. Every constraint $\la_i = \la_j$ also brings forth a factor 
$p(i,j)$, the probability for the pair to be bound. Since a bound pair
belongs by definition to the same cluster, we see the beginning of the 
geometrical mapping in which spins in a single cluster have parallel spins. To
see the full mapping, we need to repeat the procedure outlined in Eqs.~(
\ref{st1})-(\ref{st5}) for {\it all} pairs of spins. Let us consider one of the
sums into which the function $Q$ has been decomposed a step before, and 
consider a pair $(m,n)$. Two possibilities arise:
\begin{enumerate}
\item Previous constraints already determine that $\la_m = \la_n$ (for example,
there could be some $k$ for which $\la_m = \la_k$ and $\la_n = \la_k$). Then, 
necessarily, $V(m,n) = U(m,n) = v(m,n)$. Hence
\bea
\lefteqn{\sum_{\left\{\la_m \atop {\mbox{previous} \atop \mbox{ constraints}}
\right\}} \exp \left[ -\beta \sum_{(i,j)- \cdots} V(i,j)\right]} \nonumber \\
 &=& e^{-\beta v(m,n)}
\sum_{\left\{\la_m \atop {\mbox{previous} \atop \mbox{constraints}}\right\}}
\exp \left[ -\beta \sum_{(i,j)- \cdots - (m,n)} V(i,j)\right] \nonumber \\
&=& p(m,n)\, e^{-\beta v(m,n)}\sum_{\left\{\la_m \atop {\mbox{prev.} \atop 
\mbox{constr.}}\right\}} \exp \left[ -\beta \sum_{(i,j)- \cdots - (m,n)} 
V(i,j)\right] \nonumber \\
&+& q(m,n) \, e^{-\beta v(m,n)}\sum_{\left \{\la_m \atop {\mbox{prev.} \atop 
\mbox{constr.}}\right\}}\exp \left[ -\beta \sum_{(i,j)- \cdots - (m,n)} 
V(i,j)\right] \nonumber\\
\label{gen1}
\eea
where we have used $p(m,n) + q(m,n) = 1$.
\item Previous constraints do not determine that $\la_m = \la_n$. Then the 
situation is as it was for the pair $(1,2)$, and the sum will split in the 
following way
\bea
\lefteqn{\sum_{\left\{\la_m \atop \mbox{constr.}\right\}}\exp \left[ -\beta
\sum_{(i,j)- \cdots} V(i,j)\right]} \nonumber \\
\qquad &=&  p(m,n)\,e^{-\beta v(m,n)}\sum_{\left\{\la_m :\, \ldots \atop 
\la_m =\la_n \right\}} \exp \left[ -\beta \sum_{(i,j)-\cdots - (m,n)} 
V(i,j)\right]\nonumber \\
\qquad &+ & q(m,n) \,e^{-\beta v(m,n)}\sum_{\left\{\la_m :\, \dots\right\}} 
\exp \left[ -\beta \sum_{(i,j)- \cdots - (m,n)} V(i,j)\right] \nonumber \\
\label{gen2}
\eea
\end{enumerate}

From the point of view of the geometrical mapping, case (1)  means that $m$ and
$n$ already belong to the same cluster by virtue of some other particles which
link them (perhaps indirectly). Hence, it doesn't matter whether they are also
directly bound to each other [a case which contributes the factor $p(m,n)$] or
not [a case which contributes the factor $q(m,n)$]. As a result, both 
contributions appear 
without the addition of any constraint. In case (2) on the other hand, the
appearance of the factor $p(m,n)$ also requires a new constraint which now
implies that $m$ and $n$ are bound and therefore belong to the same 
cluster. The factor $q(m,n)$ requires no such constraint, since if the 
particles belong to different clusters, their spin attribution is random.

When all pairs have been covered, the set of constraints of a particular sum
specifies exactly which particles belongs to which clusters in the original
percolation model configuration. Since the expression for $Q$ contains sums 
over all possible constraints, it can be rewritten as a sum over all possible 
clusterings of the original percolation configuration. Thus, let us define
\be
P(\mbox{ conn.}) \equiv \prod_{\left\{{{\mbox{ all} \atop \mbox{bound}}
\atop {\mbox{pairs } \atop (i,j)}} \right\}} \!\!\! \! p(i,j)
\prod_{\left\{{\mbox{ all} \atop {\mbox{unbound} \atop {\mbox{pairs} \atop
(m,n)}}} \right\}} \!\!\!\! q(m,n) \; \exp\left[-\beta \sum_{i>j} v(i,j)\right]
\label{definp}
\ee
Then we can write $Q$ as
\be
Q = \sum_{\left\{ {\mbox{ all possible} \atop {\mbox{ connectivity} \atop
\mbox{ states}}}\right\}}\sum_{\left\{{{\la_m \mbox{ consistent} 
\atop\mbox{ with the }} \atop {\mbox {connectivity} \atop \mbox{state}}}
\right\}}\!\!\!\! P(\mbox{ conn.})
\label{gen3}
\ee
where the sum over all spins is consistent with the connectivity state in the
sense of the geometrical mapping, \ie, that all particles within a single 
cluster must be assigned the same spin (note, however, that not all particles 
in a cluster need be {\it directly} bound to each other).

From the probabilistic interpretation of the functions $p(i,j)$, $q(i,j)$ and 
of the usual canonical Gibbs distribution, we now see that up to a 
normalization, Eq.~(\ref{definp}) means that
\bea
P(\mbox{ conn.}) &\propto &
\mbox{ Probability density of finding a configuration} (\r_1,\ldots,\r_N)
\nonumber \\ 
& & \mbox{ such that all pairs } (i,j) \mbox{ are bound and all pairs } 
(m,n) \nonumber \\ 
& & \mbox{ are unbound.} \label{prob}
\eea
Returning now to Eq.~(\ref{conf}), and restoring the external field $h(\r)$, we
finally find
\be
Z = \frac{1}{N!} \int d1 \cdots dN \sum_{\left\{{\mbox{conn.} \atop 
\mbox{states}}\right\}}P(\mbox{ conn.}) \sum_{\{\la_m \mbox{: cl}\}}\exp\left[ 
\beta \sum_{i=1}^N h(\r_i) \p (\la_i) \right]
\label{mapz}
\ee
Here the symbol $\sum_{\{\la \mbox{: cl}\}}$ means a sum over all spin 
configurations consistent with the clustering, \ie, with the connectivity state
in the sense of the geometrical mapping.

This expression has the form of an average over all percolation configuration 
(up to nonimportant normalization factors). It is the fundamental relation
between statistical averages calculated in the percolation model and quantities
calculated in the corresponding Potts fluid. As we shall see in a moment, it
allows us to obtain percolation-related quantities by calculating properties
of the Potts fluid defined in Eq.~(\ref{map}). However, since there is no
equivalent in the percolation model to the field $h$, we need to set
this field to zero. In this case, Eq.~(\ref{mapz}) becomes
\be
Z = \frac{1}{N!} \int d1 \cdots dN \sum_{ \left\{{\mbox{conn.} \atop 
\mbox{states}}\right\}}P(\mbox{conn.}) \quad s^{N_c}
\label{mapz2}
\ee
where $N_c$ is the total number of clusters in the relevant connectivity state.
The term $s^{N_c}$ is then the number of possible assignments of spins to these
$N_c$ clusters in accordance with the geometrical mapping.

Finally we shall have to get rid of the free parameter $s$. As in the usual 
Kasteleyn-Fortuin mapping \cite{kast}, the interesting case is the limit 
$s \to 1$. The resulting expression
\be
Z_p \equiv \frac{1}{N!} \int d1 \cdots dN \sum_{\left\{{\mbox{conn.} \atop 
\mbox{states}}\right\}}P(\mbox{conn.})
\label{norm}
\ee
is just the normalization factor required for averages performed within the
percolation model. Hence,
\be
Z = Z_p \left\langle \sum_{\{\la_m \mbox{: cl}\}}
\exp\left[ \beta \sum_{i=1}^N h(\r_i) \p (\la_i) \right] \right\rangle_p
\label{basic}
\ee
where the symbol $\left\langle{\; }\right\rangle_p$ means a canonical average 
performed in the percolation model, \ie,
\be
\left\langle F (1, \ldots, N) \right\rangle_p = \frac{1}{N! Z_p} \int d1 \cdots
dN \sum_{\left\{{\mbox{conn.} \atop \mbox{states}}\right\}}P(\mbox{conn.}) \> 
F (1,\ldots,N)
\label{rel}
\ee
where $F$ is some quantity defined in the percolation model.

Similarly, we need averages performed in the Potts fluid, which will be
denoted by $\left\langle{\; }\right\rangle_s$. Given a quantity $G$ defined in 
the Potts fluid, we have
\bea
\left\langle G (1,\la_1; \ldots; N, \la_N) \right\rangle_s &=& 
\frac{1}{N! Z} \int d1 \cdots dN \sum_{\{\la_m \}} G (1,\la_1; \ldots; 
N, \la_N)\nonumber \\
& & \times\exp \left[ - \beta \sum_{i>j} V(i,j) + \beta \sum_{i =1}^N h(i) 
\p (\la_i) \right]  \nonumber \\
\label{pav}
\eea
Eq.~(\ref{gen3}) now implies that in general, for any quantity $G (1,\la_1; 
\ldots; N, \la_N)$ defined in the Potts fluid system, we have that
\be
\left\langle G (1,\la_1; \ldots; N, \la_N) \right\rangle_s = 
\left\langle  \sum_{\{\la_m \mbox{: cl}\}} G (1,\la_1; \ldots; N, \la_N)
e^ {\left[\beta \sum h(i) \p (\la_i) \right]}
\right\rangle_p
\label{fundam}
\ee
This fundamental relation allows us to translate every quantity in the Potts 
fluid into some quantity defined in the continuum percolation model.

We now turn to using Eq.~(\ref{basic}) to show that the average magnetization
in the Potts fluid is directly related to the order parameter of the 
percolation model, the percolation probability.

\section{The Potts magnetization and the percolation probability}
\setcounter{equation}{0} 

The magnetization of the Potts fluid is defined as in the usual Potts model by
\be
M = \frac{1}{\beta N (s-1)} \, \frac{\partial \ln Z}{\partial h}
\label{defm}
\ee
where $h$ is the now constant external field. From Eqs.~(\ref{basic}) and 
(\ref{fundam}), we have
\bea
M &=& \left\langle\frac{1}{N(s-1)}\sum_{i=1}^N \p (\la_i) \right\rangle_s
\nonumber \\
&=&\left\langle\frac{1}{N(s-1)}\sum_{\{\la_m \mbox{: cl}\}} \left\{
\sum_{i=1}^N \p (\la_i) \exp\left[ \beta \sum_{i=1}^N h(\r_i) \p (\la_i)\right]
\right\}\right\rangle_p
\label{mav}
\eea

In order to calculate this expression, we need to characterize more formally 
the sum over all spin states. In a given connectivity state, let us denote by 
$N_r \>(1 \le r \le N)\,$ the number of clusters containing exactly $r$ 
particles
(possibly $N_r = 0$). Also, because of their particular role, let us denote 
separately the number of spanning clusters by $N_s$ (as we take $N \to \infty$,
$N_s$ is either 0 or 1). Any connectivity state thus corresponds to a set of 
numbers $(N_1,\, N_2,\ldots,\, N_N,\,N_s)$ (other connectivity states may also 
correspond to the same set). For a given spin configuration consistent with
this clustering, let us denote by $k_r$ the number of $r$-clusters which have
been assigned the spin 1 ($k_s$ will represent the spanning clusters assigned 
the spin 1). Every spin configuration generates therefore a set 
$(k_1,\, k_2,\ldots,\, k_N, \, k_s)$. Then, the total number of spin 
configurations consistent with the given clustering is
\be
\sum_{k_1 = 0}^{N_1}\sum_{k_2= 0}^{N_2} \cdots \sum_{k_N = 0}^{N_N}
\sum_{k_s = 0}^{N_s} {k_1 \choose N_1} \cdots {k_N \choose N_N}{k_s \choose 
N_s} (s-1)^{N_1 - k_1} \cdots (s-1)^{N_s - k_s} \label{count}
\ee
The factor $(s-1)^{N_r - k_r}$ is the number of possible spin assignments to 
the $(N_r - k_r)\>$ $r$-clusters which have a spin different from 1. 

Consider now the expression $\sum_{i=1}^N \p (\la_i)$. From the definition of
$\p (\la)$, Eq.~(\ref{psi}), every $j$-cluster with a spin 1 contributes a term
$j(s-1)$ to this expression. Every $j$-cluster with a spin different from 1 
contributes $j(-1)$ to it. Therefore, for a given spin configuration,
\bea
\sum_{i=1}^N \p (\la_i) &=& (s-1) \left[ \sum_{j=1}^N j k_j + n_s k_s \right] 
- \left[ \sum_{j=1}^N j (N_j - k_j) + n_s (N_s -k_s) \right] \nonumber \\
& =& s \left[ \sum_{j=1}^{N_s} j k_j\right] - N \label{sumpsione}
\eea
where $n_s$ is the number of spins in the spanning cluster (we anticipate the 
thermodynamic limit and assume that the spanning cluster, if it exists, is 
unique), and where we have used the identity $\sum_{j=1}^{N_s} j k_j = N $. The
notation $\sum_{j=1}^{N_s} j k_j$ is shorthand for $n_s k_s + \sum_{j=1}^N j 
k_j$.

We now have that
\bea
\lefteqn{\frac{1}{N(s-1)} \sum_{\{\la_m \mbox{: cl}\}} \left\{
\exp\left[ \beta h \sum_{i=1}^N \p (\la_i) \right] \sum_{i=1}^N \p (\la_i) 
\right\}}\nonumber \\
 &=& e^{-\beta h N} \prod_{l=1}^{N_s} \Bigg\{ \sum_{k_l =0}^{N_l} 
{k_l \choose N_l} (s-1)^{N_l - k_l}\left( \ex \right)^{k_l} \nonumber \\
& & \qquad\qquad \times \left[\frac{1}{N(s-1)} \sum_{j=1}^{N_s}
\left( s j k_j \right) - N \right] \Bigg\} \label{sump2}
\eea
where $\prod\limits_{l=1}^{N_s} \equiv \prod\limits_{l=1}^{N_N} \cdots \times 
(\mbox{case}\> l= N_s)$.

It is easy to prove (\eg, by induction) the two identities
\bea
\sum_{k_l =0}^{N_l} {k_l \choose N_l} (s-1)^{N_l - k_l} \left( \ex \right)
^{k_l} &=& \left[ \ex + (s-1) \right]^{N_l}\\
\sum_{k_l =0}^{N_l} k_l {k_l \choose N_l} (s-1)^{N_l - k_l} \left( \ex 
\right)^{k_l} &=& \frac{\left[ \ex + (s-1) \right]^{N_l}}{ \ex + (s-1) } N_l 
\, \ex
\eea
Therefore,
\be
\prod_{l=1}^{N_s} \left[\sum_{k_l =0}^{N_l} {k_l \choose N_l} (s-1)^{N_l- k_l} 
\left( \ex \right)^{k_l}\right](-N) = -N \prod_{l=1}^{N-s}\left[ \ex +(s-1) 
\right]^{N_l}
\ee
and
\bea
\lefteqn{\prod_{l=1}^{N_s} \left\{\sum_{k_l =0}^{N_l} k_l {k_l \choose N_l} 
(s-1)^{N_l -k_l} \left( \ex \right)^{k_l} \sum_{j=1}^{N_s}j s \right\} } 
\nonumber \\ 
& =& \left[\prod_{{l=1 \atop l \ne j}}^{N_s} \sum_{k_l =0}
^{N_l}{k_l \choose N_l} (s-1)^{N_l -k_l} \left( \ex \right)^{k_l}\right]
\nonumber \\
&& \times \sum_{k_j =0}^{N_j} k_j {k_j \choose N_j} (s-1)^{N_j -k_j} 
\left( \exj \right)^{k_j} \nonumber \\
& =& \sum_{j=1}^{N_s} s j N_j \frac{ \exj}{ \exj + (s-1) } 
\prod_{l=1}^{N_s} \left[ \ex + (s-1) \right]^{N_l}
\label{expsi}
\eea

Combining these two equations, we finally obtain
\bea
\lefteqn{\sum_{\{\la_m \mbox{: cl}\}} \left\{\exp\left[\beta h \sum_{i=1}^N 
\p (\la_i)\right]\sum_{i=1}^N \p (\la_i)\right\}} \nonumber \\
 &=& e^{-\beta h N} \prod_{l=1}^{N_s} \left[ \ex + (s-1)\right]^{N_l}
\left\{ -N + \sum_{j =1}^{N_s} s j N_j \frac{\exj}{\exj + (s-1)} \right\}
\nonumber \\
\label{sumpsi}
\eea
Recalling that $N = \sum_{j=1}^{N_s} j N_j $ and rearranging terms, we have 
that
\bea
\lefteqn{\frac{1}{N(s-1)} \sum_{\{\la_m \mbox{: cl}\}}
\left\{ \exp \left[\beta h \sum_{i=1}^N \p (\la_i)\right]\sum_{i=1}^N 
\p (\la_i) \right\}}\nonumber \\
&=& \frac{1}{N}e^{-\beta h N} \prod_{l=1}^{N_s} \left[ \ex + (s-1)\right]^{N_l}
\left\{\sum_{j =1}^{N_s} j N_j\frac{\exj - 1}{\exj + (s-1)} \right\} \nonumber
\\
&=& \prod_{l=1}^{N_s} \left[ 1 + (s-1) e^{-s\beta l h}\right]^{N_l}\sum_{j =1}
^{N_s} \frac{j N_j}{N} \left[\frac{ 1 - \exm}{1 + (s-1)\exm} \right]
\label{sumpsitwo}
\eea
Hence
\be M = \left\langle \prod_{l=1}^{N_s} \left[ 1 + (s-1) e^{-s\beta l h}
\right]^{N_l}\sum_{j =1}^{N_s} \frac{j N_j}{N} \left[ \frac{1 - \exm}{1 + (s-1)
\exm} \right]\right\rangle_p
\label{magfin}
\ee

To obtain a quantity directly related to the percolation model, we need to take
the limit $s \to 1$, which yields
\be
M_{(s \to 1)}= \left\langle \frac{1}{N} \sum_{j=1}^{N_s} j N_j - \frac{1}{N}
\sum_{j=1}^{N_s} \exs \right\rangle_p = 1 - \left\langle \frac{1}{N}
\sum_{j=1}^{N_s} j N_j \exs \right\rangle_p
\label{magh}
\ee

Finally, we need to take the thermodynamic limit and set the external field to
0. As always, we must be careful in the order of these limits \cite{binney}. 
A broken symmetry state will only be obtained if the field remains finite while
$N \to \infty$. In the limit $N \to \infty$, the spanning cluster behaves 
differently from the finite clusters. Eq.~(\ref{magh}) can be rewritten as
\be
M =  1 - \frac{1}{N}\left\langle \sum_{j=1}^{N} j N_j \exs + \frac{n_s N_s}{N}
e^{- \beta h  n_s} \right\rangle_p
\label{magh2}
\ee
In the limit $N \to \infty$, the factors $\exp (- j \beta h)$ remain finite, 
but $\exp ( -\beta h n_s) \to 0$, because the spanning cluster then
becomes infinite. It is this property which distinguishes it from all finite 
clusters. Hence
\be
M = 1 - \lim_{N \to \infty}\left\langle \frac{1}{N}\sum_{j=1}^N j N_j \exs
\right\rangle_p   
\label{magth}
\ee

Finally,we can set $h=0$. At this stage, the limits commute and we obtain
\be
M = 1 - \lim_{N \to \infty}\left\langle \frac{1}{N}\sum_{j=1}^N j N_j \right
\rangle_p = 1 - \lim_{N \to \infty}\left\langle \frac{N - n_s}{N}\right
\rangle_p = \lim_{N \to \infty}\left\langle \frac{n_s}{N} \right\rangle_p
\label{magrel}
\ee
This is exactly the probability that a particle picked at random belongs to the
infinite cluster, which is by definition the percolation probability $P(\rho)$.
Therefore,
\be
\lim_{h \to 0} \, \lim_{N \to \infty} \, \lim_{s \to 1} \, M = P(\rho)
\ee
and the percolation order parameter is directly calculable from the Potts 
magnetization.

We can calculate the Potts susceptibility most easily from Eq.~(\ref{magth}), 
before we set the field to 0 (it is easy to see that the derivative $\partial /
\partial h$ commutes with all the operations performed up to that point). Thus,
\be
\chi = \frac{\partial M}{\partial h} = \beta \lim_{N \to \infty} \left\langle 
\frac{1}{N}\sum_{j=1}^N j^2 N_j \exs \right\rangle_p
\label{sush}
\ee
Taking now the limit $h \to 0$ and using again that $N= \sum\limits_{j=1}^{N_s}
j N_j$ gives
\be
\chi = \frac{\partial M}{\partial h} = \beta \lim_{N \to \infty} \left\langle 
\frac{\sum\limits_{j=1}^N j^2 N_j}{\sum\limits_{j=1}^{N_s} j N_j} 
\right\rangle_p
\label{sus}
\ee
For densities {\it lower} than the critical density, there is no spanning 
cluster (in the thermodynamic limit), so that
\be
\chi =\beta \lim_{N \to \infty} \left\langle 
\frac{\sum\limits_{j=1}^N j^2 N_j}{\sum\limits_{j=1}^{N} j N_j} 
\right\rangle_p = \beta S \qquad\qquad (\rho < \rho_c)
\label{sus2}
\ee
where $S$ is the average mean cluster size, by definition \cite{stauffer}.

Therefore, we can calculate the percolation probability by calculating instead 
the magnetization of a corresponding Potts fluid, and the average mean cluster
size by calculating the Potts susceptibility. 

\section{Correlation functions}
\setcounter{equation}{0} 

The Potts magnetization and susceptibility are directly related to the 
$n$-density functions of the Potts fluid, which are defined in analogy
to the $n$-density functions of a classical liquid \cite{hansen} as
\bea
\lefteqn{\rho^{(n)} (\r_1, \la_1; \r_2,\la_2; \ldots, \r_n, \la_n) =}
\nonumber \\
& &  \frac{1}{Z(N-n)!} \int d\r_{n+1} \cdots d\r_N \, \exp\left[ - \beta 
\sum_{i>j} V(i,j) - \beta \sum_{i=1}^N h(i) \p (\la_i) \right] \nonumber \\
\label{defr}
\eea
The normalization is chosen so that
\be
\sum_{\{\la_1, \ldots, \la_n\}} \int d1 \cdots dn \, \, \rho^{(n)} = s^N 
\frac{N!}{(N-n)!}   \label{normr}
\ee

The $n$-density functions can also be expressed as canonical averages,
\bea
\rho^{(1)} (\vec{x}, \sigma) &=& \left\langle \sum_{i=1}^N \delta (\r_i -
\vec{x}) \, \delta_{\la_i, \sigma} \right\rangle_s \\
\rho^{(2)} (\vec{x}, \sigma; \vec{y}, \eta) &=& \left\langle 
\sum_{i=1}^N \sum_{{j=1 \atop j \ne i}}^N\delta (\r_i -\vec{x}) \,
\delta (\r_j -\vec{y})\,\delta_{\la_i, \sigma} \, \delta_{\la_j, \eta} 
\right\rangle_s \\
\vdots && \nonumber
\eea

Now, we can rewrite
\be
\p (\la_i) = (s-1) \delta_{\la_i,1} - \sum_{\la \ne 1} \delta_{\la_i, \la}
\ee
Using the identity $\int d \vec{x} \, \delta (\r_i - \vec{x}) = 1$, we now
have 
\bea
\left\langle \sum_{i=1}^N \p (\la_i) \right\rangle_s
&=&\left\langle (s-1)\sum_{i=1}^N \int d \vec{x}\,
\delta (\r_i - \vec{x})\delta_{\la_i,1}  \right\rangle_s \nonumber \\
&-&\left\langle  \sum_{i=1}^N \int d \vec{x}\, \delta (\r_i - \vec{x})
\sum_{\la \ne 1} \delta_{\la_i, \la} \right\rangle_s \nonumber \\
&=& \int d \vec{x} \left[ (s-1) \rho^{(1)} ( \vec{x},
\sigma = 1) - \sum_{\sigma \ne 1} \rho^{(1)} (\vec{x}, \sigma) \right]
\label{mrho}
\eea
Since the field $h$ preserves the symmetry between all the spins $\la \ne 1$, 
$\rho^{(1)} (\vec{x}, \sigma)$ must be the same for all $\sigma \ne 1$. 
Hence,
\be
M = \left\langle \frac{1}{N(s-1)} \sum_{i=1}^N \p (\la_i) \right\rangle_s =
\frac{1}{N} \int d \vec{x} \left[ \rho^{(1)} ( \vec{x},1) -
\rho^{(1)} (\vec{x}, \alpha) \right]
\ee
where $\alpha$ denotes some (arbitrary) spin value other than $1$. From the 
normalization, Eq.~(\ref{normr}),it is now obvious that
\bea
\int d \x \, \rho^{(1)} ( \vec{x}, 1) &=& n_1 \equiv
\mbox{ number of spins in state } 1 \nonumber \\
\int d \x \, \rho^{(1)} ( \vec{x},\alpha) &=& n \equiv
\mbox{ number of spins in any state } \alpha \ne 1
\eea
Hence
\be
M = \frac{1}{N} \left( n_1 - n \right)
\ee
is just the excess density of spins in the state $1$ over the density of spins
in any other state.

From Eq.~(\ref{mav}), and the definition $\chi = \partial M / \partial h$, 
the susceptibility is
\be
\chi = \left\langle \frac{\beta}{ N(s-1)} \sum_{i\ne j}\p(\la_i)\p (\la_j)
\right\rangle_s + \left\langle \frac{\beta}{ N(s-1)} \sum_{i=1}^N \p^2(\la_i)
\right\rangle_s  \label{chis}
\ee
Repeating the steps leading to Eq.~(\ref{mrho}), we have
\be
\left\langle \frac{\beta}{ N(s-1)} \sum_{i=1}^N \p^2(\la_i)\right\rangle_s = 
\frac{\beta}{N(s-1)} \int d \vec{x} \sum_{\sigma}\p^2 (\sigma) \, \rho^{(1)} 
( \vec{x},\sigma)
\ee
where the sum over $\sigma$ runs over all the $s$ possible spin states.

On the other hand,
\be
\sum_{i\ne j}\p(\la_i)\p (\la_j) = \sum_{\alpha} \sum_{\gamma}\p(\alpha)
\p (\gamma) \sum_{i \ne j} \delta_{\la_i, \alpha} \delta_{\la_j, \gamma}
\ee
where, again, the sum over $\alpha$ and $\gamma$ runs over the $s$ spin states.
Hence, by a derivation similar to the one used for Eq.~(\ref{mrho}), we have
\bea
\lefteqn{\left\langle \sum_{i\ne j}\p(\la_i)\p (\la_j)\right
\rangle_s} \nonumber \\
&=& \left\langle \sum_{\alpha} \sum_{\gamma}
\p(\alpha)\p (\gamma) \sum_{i \ne j} \int d \x \, d \y \,\,\delta (\r_i - \x)
\, \delta (\r_j - \y)\, \delta_{\la_i,\alpha}\, \delta_{\la_j,\gamma}
\right\rangle_s \nonumber \\
&=& \int d \x \, d \y \left[ \sum_{\alpha} 
\sum_{\gamma}\p(\alpha)\p (\gamma)\, \rho^{(2)} ( \vec{x},\alpha;\vec{y}, 
\gamma) \right]
\label{chirho}
\eea
Therefore
\bea
\chi &=& \frac{\beta}{N(s-1)} \int d \x \, d \y \left[\sum_{\alpha} 
\sum_{\gamma}\p(\alpha)\p (\gamma)\, \rho^{(2)} ( \vec{x},\alpha;\vec{y}, 
\gamma) \right] \nonumber \\
&+& \frac{\beta}{N(s-1)} \int d \x \sum_{\alpha}\p^2 
(\alpha) \, \rho^{(1)} ( \x,\alpha)
\eea

let us now relate this expression to the percolation picture, \ie, to the limit
$\, s \to 1$, when $\chi \to \beta S\,$ for densities $\rho < \rho_c$. For this
range of densities, the symmetry of the system is completely unbroken and
$\rho^{(1)} (\vec{x}, \alpha)$ is independent of $\alpha$. From the 
normalization Eq.~(\ref{normr}), it follows that
\be
\int d \x \, \rho^{(1)} ( \vec{x},\alpha) = \frac{1}{s} \int d 
\vec{x} \sum_{\alpha} \rho^{(1)} ( \vec{x},\alpha) = N
\ee
Now $ \sum_{\alpha} \p^2 (\alpha) = (s-1)^2 + (s-1) (-1)^2$, so that
\be
\lim_{s \to 1} \frac{1}{s-1} \sum_{\alpha} \p^2 (\alpha) = 1
\ee
and therefore
\be
\frac{\beta}{N(s-1)} \int d \vec{x} \sum_{\alpha}\p^2 (\alpha) \, \rho^{(1)} 
( \vec{x},\alpha) = \beta
\ee
Hence, when $s \to 1$,
\be
\frac{\chi}{\beta} \to S = 1 + \lim_{s \to 1} \, \frac{1}{N(s-1)} \int 
d \vec{x} \, d \vec{y} \left[\sum_{\alpha} \sum_{\gamma}\p(\alpha)\p 
(\gamma)\rho^{(2)} ( \vec{x},\alpha;\vec{y}, \gamma) \right] 
\label{Spotts}
\ee

Let us now recalculate this quantity by using the general connection between 
averages in the Potts fluid and in the percolation model, Eq.~(\ref{fundam}). 
In particular,we have
\be
\left\langle \frac{\beta}{ N(s-1)} \sum_{i\ne j}\p(\la_i)\p (\la_j)
\right\rangle_s = \frac{\beta}{ N(s-1)} \sum_{i\ne j} 
\left\langle \sum_{\{\la_m:\, cl.\}} \p(\la_i)\p (\la_j)\right\rangle_p 
\ee
where we have already set $h = 0$, which is allowed since in the range $\rho <
\rho_c$, the symmetry is unbroken anyway. Let us now calculate the average on 
the r.h.s separately when $\la_i$ and $\la_j$ belong to the same cluster and 
when they do not. To this end, define a function $\Omega (i,j)$ as
\be
\Omega (i,j)= \left\{ \begin{array}{r @{\qquad}l}
   1 & \mbox{if \quad}i , j \mbox{\quad belong to the same cluster}\\
   0 & \mbox{otherwise} 
   \end{array} \right.    \label {omega}
\ee
Then, from Eq.~(\ref{rel}),
\bea
\lefteqn{\left\langle \sum_{\{\la_m:\, cl.\}}\p(\la_i)\p (\la_j)\right
\rangle_p} \nonumber \\
 &=& \frac{1}{N! Z_p} \int d 1 \cdots d N \sum_{\left\{{\mbox{conn.} \atop 
\mbox{states}}\right\}}P (\mbox{conn.})\sum_{\{\la_m:\, cl.\}}\p(\la_i)
\p (\la_j)\Omega (i,j) \nonumber \\
&+& \frac{1}{N! Z_p} \int d 1 \cdots d N \sum_{\left\{{\mbox{conn.} \atop 
\mbox{states}}\right\}} P(\mbox{conn.}) \sum_{\{\la_m:\, cl.\}}\p(\la_i)
\p (\la_j) \left[ 1 - \Omega (i,j)\right] \nonumber \\
\label{omeg2}
\eea
The first sum contributes only if $\la_i, \la_j$ belong to the same cluster, 
while the second contributes only if they belong to separate clusters. Because
of the geometrical mapping, $\p(\la_i)= \p (\la_j)$ in the first sum. Also, in
the sum over $\{ \la_m: \, cl\}$, every cluster other than the one containing
$\la_i$ and $\la_j$ contributes a factor $s$, the number of possible spin
assignments. The cluster containing $\la_i$ and $\la_j$ , on the other hand, 
contributes a factor $(s-1)^2$ if $\la_i=\la_j=1$ and $(-1)^2$ for the $(s-1)$ 
other possible choices for $\la_i = \la_j$. Therefore,
\be
\sum_{\{\la_m:\, cl.\}}\p(\la_i)\p (\la_j)\Omega (i,j) = s^{N_c -1}
\left[(s-1)^2 + (s-1)(-1)^2 \right]\Omega (i,j)
\ee
where $N_c$ is the total number of cluster in the configuration. Hence,
\bea
\lefteqn{\frac{\beta}{Z_p N (s-1) N!} \int d 1 \cdots d N \sum_{\left\{
{\mbox{conn.} \atop \mbox{states}}\right\}}P(\mbox{conn.}) 
\sum_{\{\la_m:\, cl.\}}\p(\la_i)\p (\la_j) \Omega (i,j)} \nonumber \\
\quad &=& \frac{\beta}{N! Z_p N} \int d 1 \cdots d N \sum_{\left\{{\mbox{conn.}
\atop \mbox{states}}\right\}}P(\mbox{conn.})s^{N_c -1}(s-1+1)\Omega (i,j)
\nonumber \\
\quad &=& \frac{\beta}{N} \left\langle s^{N_c}\Omega(\la_i,\la_j)\right
\rangle_p
\label{omeg}
\eea

The second sum in the r.h.s of Eq.~(\ref{omeg2}) contributes only if 
$\la_i, \la_j$ belong to different clusters. The cluster containing $\la_i$ 
contributes a factor $(s-1)$ if $\la_i = 1$, and a factor $(-1)$ in all the 
other $(s-1)$ cases. The same holds for the cluster containing $\la_j$. The 
$N_c - 2$ remaining clusters contribute each a factor $s$. Hence,
\bea
\lefteqn{\sum_{\{\la_m:\, cl.\}}\p(\la_i)\p (\la_j)\left[ 1 - \Omega (i,j)
\right]} \nonumber \\
 &=& s^{N_c -2}\left[(s-1) + (s-1)(-1)\right]^2 \left[1 - \Omega (i,j)\right] 
= 0  \label{omegnot}
\eea
We can now substitute the results Eqs.~(\ref{omeg}) and (\ref{omegnot}) into
Eq.~(\ref{chis}), then take again the limit $s \to 1$. Repeating the steps
leading to Eq.~(\ref{Spotts}), we end up this time with
\be
\frac{\chi}{\beta} \to S = 1 + \frac{1}{N} \int d \x \, d \y \left \langle 
\sum_{i \ne j}\Omega (i,j)\delta (\r_i - \x)\delta (\r_j - \y) \right\rangle_p
\label{Spc}
\ee

We define now the function
\be
\gd (\vec{x}, \vec{y}) \equiv \frac{1}{\rho(\vec{x})\rho(\vec{y})}  \left
\langle \sum_{ i \ne j}\Omega (i,j)\delta ( \r_i - \vec{x})
\delta ( \r_j - \vec{y}) \right\rangle_p 
\label{defgd}
\ee
to be the {\it pair connectedness} function [$\rho(\vec{x})$ is the density at
position $\vec{x}$]. From the definition of $\Omega (i,j)$, the 
meaning of $\gd$ is 
\bea
\rho(\vec{x})\rho(\vec{y}) \, \gd (\vec{x}, \vec{y})\, d \vec{x} d \vec{y} &=& 
\mbox{Probability of finding two particles in regions } \nonumber \\
& & d \vec{x} \mbox{ and } d \vec{y} \mbox{ around the positions }
\vec{x} \mbox{ and } \vec{y} \mbox{, such} \nonumber \\
& & \mbox{that they both belong to the same cluster.} \nonumber\\
\eea
which shows the pair-connectedness to be a generalization of the corresponding
function in lattice percolation \cite{stauffer}. Hence,
\be
S = 1 + \frac{1}{N} \int d \x \, d \y \,\,\rho(\x)\rho(\y) \, \gd (\x, \y)
\label{Sone}
\ee
Usually, the system is translationally invariant, so that $\gd (\x,\y) = 
\gd (\x- \y)$ and $\rho(\x)= \rho(\y) = \rho$. Then
\be
S = 1+\frac{V}{N} \rho^2 \int d \r \, \gd (\r) = 1 + \rho\int d \r \, \gd (\r)
\label{Stwo}
\ee
Comparing Eq.~(\ref{Sone}) with Eq.~(\ref{Spotts}), we finally obtain the
important relationship
\be
\gd (\vec{x}, \vec{y}) =  \lim_{s \to 1} \, \frac{1}{(s-1)\rho^{(1)}(\x)\,
\rho^{(1)}(\y)}
\sum_{\alpha} \sum_{\gamma}\p(\alpha)\p (\gamma)\rho^{(2)} 
( \vec{x},\alpha;\vec{y}, \gamma) 
\label{gdconn}
\ee
Let us now introduce the spin pair-correlation function, defined as
\be
g_s^{(2)} (\vec{x},\alpha; \vec{y}, \gamma) \equiv 
\frac{1}{\rho^{(1)}(\x)\,\rho^{(1)}(\y)} \rho^{(2)}( \vec{x},\alpha;\vec{y}, 
\gamma) \label{pcorr}
\ee
which tends to $1$ when $| \vec{x} - \vec{y}| \to \infty $. We can now rewrite
Eq.~(\ref{gdconn}) as
\be
\gd (\vec{x}, \vec{y}) = \lim_{s \to 1} \, \frac{1}{s-1}
\sum_{\alpha} \sum_{\gamma}\p(\alpha)\p (\gamma) \, g_s^{(2)}( \x,\alpha;\y, 
\gamma) 
\label{gdgp}
\ee
or
\bea
\gd (\vec{x}, \vec{y}) &=& \lim_{s \to 1} \frac{1}{s-1}\sum_{\alpha}\p ^2
(\alpha)g_s^{(2)}( \vec{x},\alpha;\vec{y}, \alpha) \nonumber \\
&+& \lim_{s \to 1} \frac{1}
{s-1}\sum_{\alpha \ne \gamma}\p(\alpha)\p (\gamma) g_s^{(2)}( \vec{x},\alpha;
\vec{y}, \gamma) 
\label{76}
\eea
We now have
\be
\sum_{\alpha}\p ^2(\alpha)g_s^{(2)}( \vec{x},\alpha;\vec{y}, \alpha) = (s-1)^2
g_s^{(2)}( \vec{x},1;\vec{y}, 1) + (s-1) (-1)^2
g_s^{(2)}( \vec{x},\sigma;\vec{y}, \sigma) 
\ee
where $\sigma$ is any value of the spin different from $1$. As a result,
\be
\lim_{s \to 1} \, \frac{1}{s-1}\sum_{\alpha}\p ^2(\alpha) \,
g_s^{(2)}( \vec{x},\alpha;\vec{y}, \alpha) = \lim_{s \to 1} \,
g_s^{(2)}( \vec{x},\sigma ;\vec{y}, \sigma) 
\label{78}
\ee
where $\sigma \ne 1$. 

Similarly,
\bea
\sum_{\alpha \ne \gamma}\p(\alpha)\p (\gamma) \, g_s^{(2)}( \vec{x},\alpha;
\vec{y}, \gamma) &=& (s-1)\sum_{\gamma \ne 1}\p (\gamma)\, g_s^{(2)}
( \vec{x},1;\vec{y}, \gamma) \nonumber \\
&+& (s-1)\sum_{\alpha \ne 1}\p (\alpha) \,
g_s^{(2)}( \vec{x},\alpha ;\vec{y},1) \nonumber \\
&-& \sum_{\alpha \ne 1}\sum_{{\gamma \ne \alpha} \atop {\gamma \ne 1}}
\p(\alpha)\p (\gamma)\, g_s^{(2)}( \vec{x},\alpha;\vec{y}, \gamma) \nonumber \\
&=& (s-1)^2 \left[ g_s^{(2)}(\vec{x},1;\vec{y}, \sigma) + g_s^{(2)}
( \vec{x},\sigma;\vec{y}, 1) \right] \nonumber \\
&+& (s-1)(s-2) \, g_s^{(2)}( \vec{x},\sigma;\vec{y}, \eta)
\eea
where $\sigma \ne\eta$ are any values of the spin which are both different
from $1$. As a result,
\be
\lim_{s \to 1} \, \frac{1}{s-1}\sum_{\alpha \ne \gamma}\p(\alpha)\p (\gamma)\,
g_s^{(2)}( \vec{x},\alpha; \vec{y}, \gamma) = - \lim_{s \to 1}\,  g_s^{(2)}
( \vec{x},\sigma;\vec{y}, \eta)
\label{80}
\ee
Substituting Eqs.~(\ref{78}) and (\ref{80}) into Eq.~(\ref{76}), we have,
finally, that
\be
\gd (\vec{x}, \vec{y}) = \lim_{s \to 1} \left[ g_s^{(2)}( \vec{x},\sigma ;
\vec{y}, \sigma) - g_s^{(2)}(\vec{x},\sigma;\vec{y}, \eta) \right] \qquad 
(\rho < \rho_c )
\label{gdagf}
\ee
where $\sigma, \eta \ne 1$ and $\sigma \ne \eta$.

This last equation is easily understood in terms of the geometrical mapping. If
two particles $\vec{x}$ and $\vec{y}$ belong to the same cluster, they must
have the same spin, say a spin $\sigma$. Hence $\gd (\vec{x}, \vec{y})$ must
be contained in $g_s^{(2)}( \vec{x},\sigma ;\vec{y}, \sigma)$. However,
$g_s^{(2)}( \vec{x},\sigma ;\vec{y}, \sigma)$ also 
includes the case where the two particles have the same spin but belong to
different clusters. This happens if the two clusters have been assigned, by
chance, the same overall spin. However, since such spin assignment is random,
the probability of the two particles having the spins $\sigma$ and $\sigma$
(identical) is exactly the same as their having spins $\sigma$ and $\eta$, 
where now $\sigma \ne \eta$ (note that such an assignment automatically 
necessitates that the particles belong to different clusters). Hence $g_s^{(2)}
(\vec{x},\sigma ;\vec{y}, \sigma)$ exceeds $\gd (\vec{x}, \vec{y})$ by 
precisely $g_s^{(2)}( \vec{x},\sigma ;\vec{y}, \eta)$. This is the meaning of
Eq.~(\ref{gdagf}). It allows us to calculate the pair-connectedness by working 
out the spin pair-correlations and then taking the limit $s \to 1$. This
completes the relation between the percolative quantities and those of the 
Potts fluid.

\section{Conclusion}
\setcounter{equation}{0} 

This paper has focused on the formal basis of continuum percolation theory.
It has provided a non-perturbative definition of the fundamental quantities
of the theory as well as showing formally how the binding criterion $ p(\r)$ 
and the interaction $v(\r)$ enter them. We have seen that the quantities of 
interest in continuum percolation can be obtained from the $s \to 1$ limit of
the Potts fluid. Specifically, the magnetization and the susceptibility become
in this limit the percolation probability and the mean cluster size, 
respectively, while the pair-connectedness is, in this limit, the difference
between two Potts pair-correlation functions. 

The advantage of this mapping is that the Potts fluid has a Hamiltonian 
formulation. This is the key to applying the techniques of equilibrium 
statistical mechanics and phase transitions to the problem of continuum 
percolation, a task which will be undertaken in future papers in this series. 
The line of attack is always to perform all calculations or theoretical 
derivations within the Potts fluid model, then to take the limit $s \to 1$ and
thus obtain the corresponding values or expressions in the percolation system.

The first such technique that one would think to apply to any problem of phase
transition is mean field theory. The Hamiltonian formulation of the Potts fluid
allows a mean field approximation to be defined, while it would be far from 
obvious how to do this directly for the continuum percolation system. Mean 
field theory turns out to be non-trivial to derive because of the continuum
nature of the system and the presence of interactions. It is the subject of the
next paper.

\end{document}